\documentclass{revtex4}

\usepackage{graphicx}
\def\bit{\begin{itemize}}
\def\eit{\end{itemize}}

\def\what{\widehat}
\def\fbi{~{\rm fb}^{-1}}

\def\sig{\sigma}
\def\gamhsmtot{\Gamma_{\hsm}^{\rm tot}}
\def\br{{\rm BF}}
\def\gamhtot{\Gamma_{\h}^{\rm tot}}

\def\hl{h^0}
\def\ha{A^0}
\def\hh{H^0}

\def\mha{m_{\ha}}
\def\mhh{m_{\hh}}

\def\h{h}
\def\mh{m_{\h}}
\def\epem{e^+e^-}
\def\mupmum{\mu^+\mu^-}
\def\anti{\overline}
\def\tanb{\tan\beta}
\def\what{\widehat}
\def\rts{\sqrt s}
\def\hsm{h_{SM}}
\def\mhsm{m_{\hsm}}

\def\gev{~{\rm GeV}}
\def\tev{~{\rm TeV}}

\def\srts{\sigma_{\!\!\!\sqrt s}^{\vphantom y}}
\def\lsim{\alt}
\def\gsim{\agt}
\def\beq{\begin{equation}}
\def\eeq{\end{equation}}
\def\bea{\begin{eqnarray}}
\def\eea{\end{eqnarray}}

\begin{document}
\bibliographystyle{revtex}

\hspace*{5.5in} {\bf IUHET-446}

\title{Higgs Bosons at Muon Colliders\footnote{Submitted to the Proceedings
of ``The Future of Particle Physics'', Snowmass 2001, P1 group. An expanded
version of the Physics of Higgs Factories has been submitted to the E1 
Group\cite{e1group}.}}



\author{M.~S.~Berger}
\email[]{berger@indiana.edu}
\affiliation{Physics Department, Indiana University, Bloomington, IN 47405}


\date{\today}

\begin{abstract}
We review the role of a muon collider in the study of Higgs bosons 
via production in the $s$-channel. Very precise measurements of a Standard 
Model-like Higgs boson mass and total width can be performed, and may 
lead to a discrimination between a Standard Model Higgs boson
and the light Higgs boson of the minimal supersymmetric theory.
The heavier Higgs bosons from a supersymmetric theory or an exotic 
Higgs sector can be studied in the $s$-channel.
A muon collider may play a crucial role in separating the overlapping 
signals for two heavy nearly degenerate Higgs bosons, and may play an 
important role in precision tests of radiative corrections in the Higgs
sector. The measurements at a muon collider will be complementary to 
the Higgs studies at the Large Hadron Collider and at an
electron-positron Linear Collider.
\end{abstract}

\maketitle


\section{Introduction}

\begin{figure}
\includegraphics{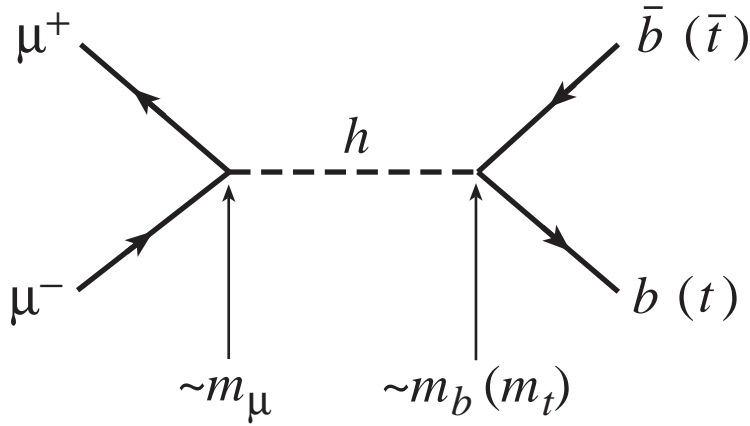}
\caption{Feynman diagram for $s$-channel production of a Higgs boson.}
\label{feyn-diag}
\end{figure}

Interest has grown rapidly in muon colliders in the last several years as 
it became clear that the technological challenges might not be 
insurmountable\cite{Ankenbrandt:1999as}. 
Muon colliders are of interest to particle physics exploration
for a number of reasons: a) the absence of significant bremsstrahlung allows
one to contemplate circular accelerators of much higher energy than is 
possible with $e^+e^-$ machines, b) the coupling of Higgs bosons is 
proportional to particle mass (see Fig.~(\ref{feyn-diag})), and 
hence there is the possibility that
Higgs bosons can be produced in reasonable numbers in the 
$s$-channel\cite{Barger:1997jm,Barger:1995hr}, 
c) there are regions of 
parameter space for which it will be impossible for either the Large Hadron 
Collider (LHC) or a Linear Collider (LC)
to discover the heavier Higgs bosons of supersymmetry or,
in the case of a general two-Higgs-doublet or more extended model,
Higgs bosons of any mass with small or zero $VV$ coupling,
d) the neutrinos from the decays of muons can be used as a source for 
a neutrino factory\cite{Ayres:1999ug,Holtkamp:2000xn}. 

The large mass of the muon in comparison to that of the electron 
results in a number of advantageous features of a muon collider. 
The beam energy spreads of a muon collider can be very small,
making them useful for studying narrow resonances like the SM Higgs 
boson. In addition, there is little bremsstrahlung, 
and the beam energy can be tuned to one part
in a million through {\it in situ} 
spin-rotation measurements\cite{Raja:1998ip}.

High rates of Higgs production at $\epem$ colliders rely on
substantial $VV$ Higgs coupling for the Higgs-strahlung process
$Z+$Higgs or for the $WW$ fusion process $WW\to$Higgs ($WW$ fusion).
In contrast, a $\mupmum$ collider can provide a factory for producing
a Higgs boson with little or no $VV$ coupling so long as it
has SM-like (or enhanced) $\mupmum$ couplings. Important examples of
this last form of Higgs boson are the heavy neutral Higgs bosons $\hh$ and 
$\ha$ of the Minimal Supersymmetric Standard Model (MSSM).

If the a light ($\lsim 130$~GeV) Higgs boson exists,
then both $\epem$ and $\mupmum$ colliders will be valuable;
the Higgs boson would have been discovered at a previous 
higher energy collider (possibly a muon collider
running at high energy), and then the Higgs factory
would be built with a center-of-mass energy 
precisely tuned to the Higgs boson mass.
The most likely scenario is that the Higgs boson 
is discovered at the LHC via gluon fusion
($gg\to H$) or perhaps 
earlier at the Tevatron via associated production 
($q\bar{q}\to WH, t\overline{t}H$), and its mass is determined to an 
accuracy of about 100~MeV. If a linear collider has also observed the Higgs
via the Higgs-strahlung process ($e^+e^-\to ZH$), one might know the Higgs 
boson mass to better than 50~MeV with an integrated luminosity of 
$500$~fb$^{-1}$.
The muon collider would be optimized to run at $\sqrt{s}\approx m_H$, and this
center-of-mass energy would be varied over a narrow range
so as to scan over the Higgs resonance. 

\section{SM-Like Higgs Bosons}

The production of a Higgs boson (generically denoted $\h$)
in the $s$-channel with interesting rates is  
a unique feature of a muon collider \cite{Barger:1997jm,Barger:1995hr}. 
The resonance cross section is
\begin{equation}
\sigma_h(\sqrt s) = {4\pi \Gamma(h\to\mu\bar\mu) \, \Gamma(h\to X)\over
\left( s - m_h^2\right)^2 + m_h^2 \left(\Gamma_{\rm tot}^h \right)^2}\,.
\label{rawsigform}
\end{equation}
In practice, however, there is a Gaussian spread ($\srts$) to
the center-of-mass energy and one must compute the
effective $s$-channel Higgs cross section after convolution 
assuming some given central value of $\rts$:
\bea
\anti\sigma_h(\sqrt s) & =& {1\over \sqrt{2\pi}\,\srts} \; \int \sigma_h  
(\sqrt{\what s}) \; \exp\left[ -\left( \sqrt{\what s} - \sqrt s\right)^2 \over  
2\sigma_{\sqrt s}^2 \right] d \sqrt{\what s}~~~
\stackrel{\rts=\mh}{\simeq}  ~~~ {4\pi\over m_h^2} \; {\br(h\to\mu\bar\mu) \,
\br(h\to X) \over \left[ 1 + {8\over\pi} \left(\srts\over\gamhtot 
\right)^2 \right]^{1/2}} \,.
\label{sigform}
\eea
\begin{figure}
\centering\leavevmode
\includegraphics[width=5in]{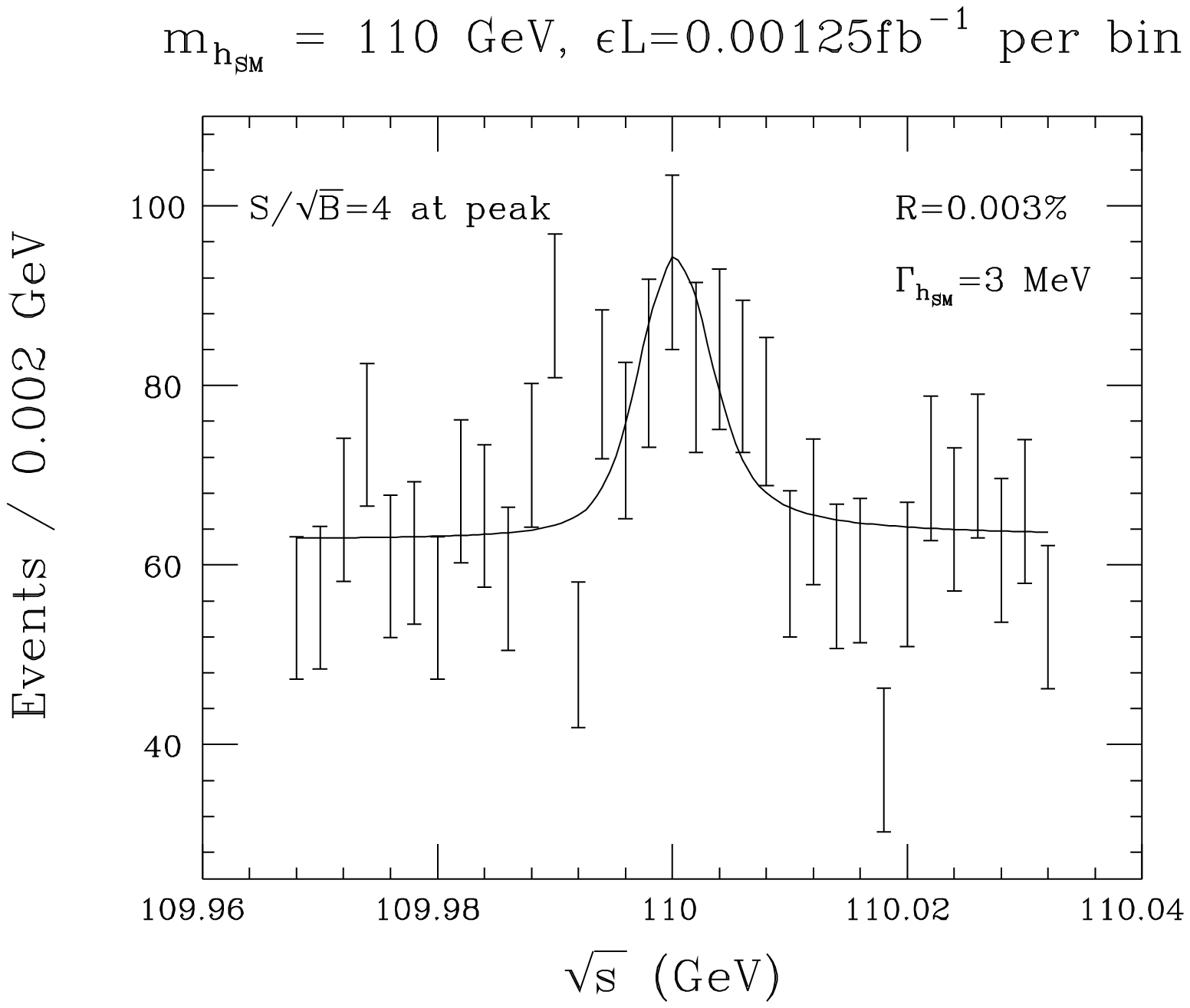}
\caption{
Number of events and statistical errors in the $b\overline{b}$
final state as a function
of $\protect\rts$ in the vicinity of $\mhsm=110\gev$,
assuming $R=0.003\%$,
and $\epsilon L=0.00125$~fb$^{-1}$ at each data point.
\label{mhsmscan}}
\end{figure}
It is convenient to express $\srts$ in 
terms of the root-mean-square (rms) Gaussian spread
of the energy of an individual beam, $R$: 
\begin{equation}
\srts = (2{\rm~MeV}) \left( R\over 0.003\%\right) \left(\sqrt s\over  
100\rm~GeV\right) \,.
\end{equation}
It is clear from Eq.~(\ref{rawsigform}) that a
resolution $\srts \lsim \gamhtot$ is needed to be
sensitive to the Higgs width. Furthermore, Eq.~(\ref{sigform}) indicates 
that $\br(\h\to \mu\anti\mu)$ must not be extremely suppressed for there to
be large event rates for Higgs production. 
The width of a light SM-like Higgs is very small (e.g. a few MeV
for $\mhsm\sim 110\gev$), implying the need for $R$
values as small as $\sim 0.003\%$ for studying a light SM-like $\h$.
In addition to the very small beam energy spread, one must also be able 
to determine very accurately the beam energy to perform a scan over such 
a narrow resonance. This can be accomplished utilizing the 
spin precession of the muon noted above. A sample scan is illustrated in 
Fig.~\ref{mhsmscan} for a $\mhsm=110\gev$ SM Higgs boson.

\begin{figure}
\centering\leavevmode
\includegraphics[width=\textwidth]{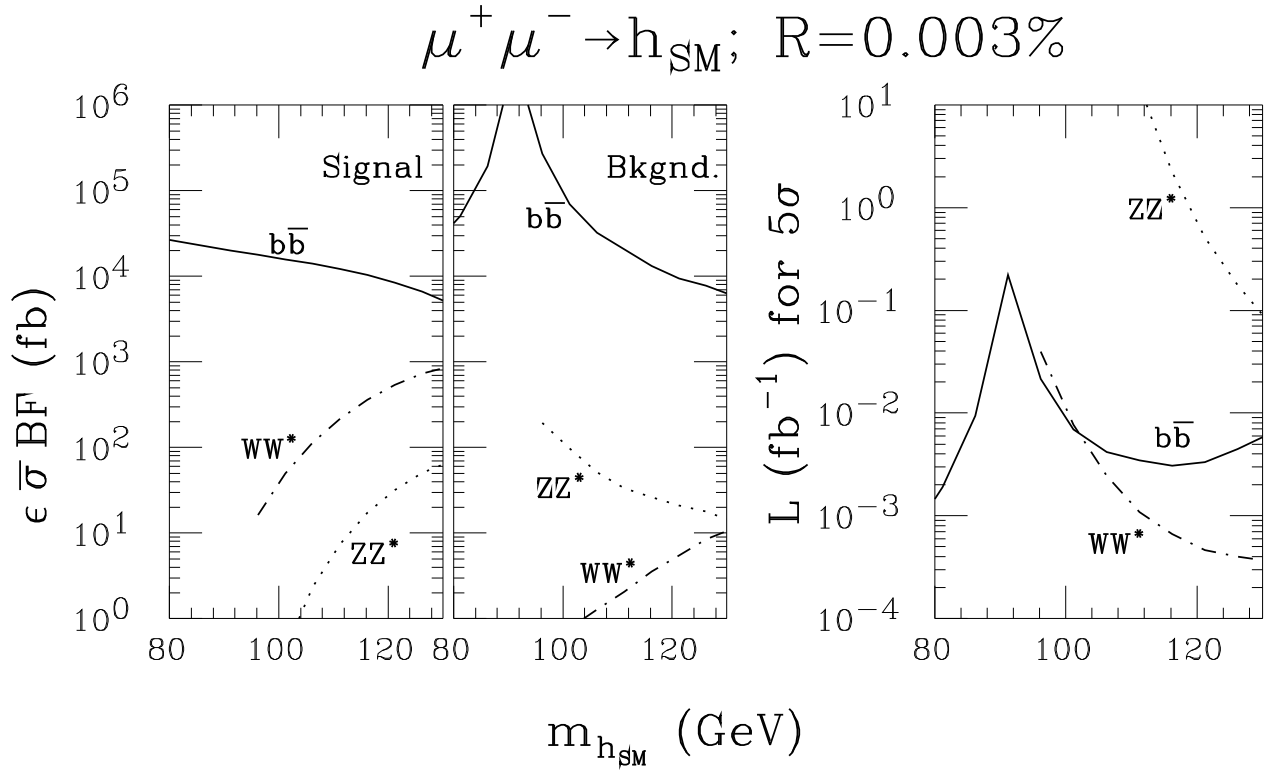}
\caption[The SM Higgs cross sections and backgrounds in $b\bar b,\ WW^*$  
and $ZZ^*$. ]{The SM Higgs cross sections and backgrounds in $b\bar b,\ WW^*$  
and $ZZ^*$. Also shown is the luminosity needed for a 5~standard deviation  
detection in $b\bar b$. From Ref.~\cite{Barger:1997jm}. For a SM-like $\h$, 
at $\sqrt s = \mh \approx 115$~GeV, the $b\bar b$ final state rates are
$\approx 10^4\:\:{\rm events\:\:\times L}(fb^{-1})$ for both the signal and
the background.
\label{sm-higgs}}
\end{figure}

The SM Higgs cross sections and backgrounds as well as the integrated 
luminosity required for a $5\sigma $ signal are shown 
in Fig.~\ref{sm-higgs} for $R=0.003\%$ and
$\mhsm$ values such that the dominant decay mode is $b\overline{b}$.
The significance of the signal is impacted by two physical processes:
1) For a Higgs mass near the $Z$-pole there is a significant background
from $\mu^+\mu^-\to Z\to b\overline{b}$. However
the most recent experimental results from LEP
have pushed the SM Higgs mass bound well above $91$~GeV. 
2) For a Higgs mass $\gsim 130$~GeV, the Higgs width $\gamhtot$ 
becomes much larger as
the $WW^\star$ decay channel opens up. 

The Higgs bosons in supersymmetric models are in general detectable at muon 
colliders.  If the masses of the supersymmetric particles are large, the
Higgs sector typically exhibits decoupling behavior in which the 
lightest supersymmetric Higgs boson $\hl$ will
be very similar to the $\hsm$ when the other Higgs bosons are
heavy, and the $\hl$ rates will be very similar to $\hsm$ rates.
On the other hand, the heavier Higgs bosons in a typical supersymmetric model
decouple from pairs of gauge bosons $VV$ at large mass  and remain reasonably
narrow ($<1$~GeV unless the $t\overline{t}$ decay mode is open). 
As a result, their $s$-channel production rates 
remain large, and a muon collider can avoid the 
production channels that
depend on a sizable coupling to gauge bosons.

What can a muon collider add to the LHC and LC? The LHC and quite likely a
linear collider will be available already, 
and the Higgs boson will be detected and 
some of its properties determined before a muon collider will become 
operational. Current expectations for 
the luminosity at an LC are 500~fb$^{-1}$ over 1-2 years. This yields a 
SM Higgs boson production rate of greater than $10^4$ per year in the  
process $e^+e^-\to Z\h$. Therefore the latest estimates of the 
luminosity at a linear collider yield numbers of Higgs bosons that are
comparable to what will be available at a muon collider/Higgs factory with 
its more modest integrated luminosity (expected with the current machine 
parameters) of the order of one inverse femtobarn.
A linear collider with such high luminosity can certainly perform quite 
accurate measurements of certain Higgs parameters such as the Higgs mass, 
couplings to gauge bosons, couplings to heavy quarks, 
etc.\cite{Battaglia:2000jb}.

The $s$-channel production process allows one to determine the mass, 
total width, and the cross sections
$\overline \sig_h(\mupmum\to\h\to X)$ 
for several final states $X$ 
to very high precision. The Higgs mass, total width and the cross sections 
can be used to constrain the parameters of the Higgs sector. 
For example, in the MSSM their precise values will
constrain the Higgs sector parameters
$\mha$ and $\tanb$ (where $\tanb$ is 
the ratio of the two vacuum expectation values (vevs) of the 
two Higgs doublets of the MSSM). The main question is whether these
constraints will be a valuable addition to LHC and LC constraints.

Precise measurements of the couplings of the Higgs boson to the Standard 
Model particles are important tests of the mass generation mechanism.
In the Standard Model with one Higgs doublet, this coupling is proportional 
to the particle mass. In the more general case there can be mixing angles
present in the couplings. Precision measurements of the couplings can 
distinguish the Standard Model Higgs boson from the SM-like
Higgs boson typically present in a more general model. If 
deviations are found, their magnitude can be extremely crucial
for constraining the parameters of the more general Higgs sector. In
particular, it might be possible to estimate the masses of the
other Higgs bosons of the extended Higgs sector, thereby allowing
a more focused search for them.

The precision possible at a muon collider
for measuring $\mh$ and $\gamhtot$ of
a SM-like $\h$ with $\mh\sim 110\gev$ are $1-3\times 10^{-6}$ and $0.2$ 
respectively.
To achieve these accuracies, one first determines the Higgs mass 
to about 1~MeV by the preliminary scan
illustrated in Fig.~\ref{mhsmscan}. Then, a dedicated
three-point fine scan\cite{Barger:1997jm} near the resonance peak
using $L\sim 0.2\fbi$
of integrated luminosity (corresponding to a few years of operation)
would be performed.
For a SM Higgs boson with a mass sufficiently below the $WW^\star$ 
threshold, the Higgs total width is very small (of order several MeV), and the 
only process where it can be measured {\it directly} is in the $s$-channel
at a muon collider. An accurate measurement of $\gamhtot$ would be
a very valuable input for precision tests of the Higgs sector.
In particular, since all the couplings 
of the Standard Model $\hsm$ are known, $\gamhsmtot$
is precisely predicted. Therefore, the precise determination of 
$\gamhtot$ obtained by this scan
would be an important test of the Standard Model, and any deviation
would be evidence for a nonstandard Higgs sector (or other new physics). 
 
Other interesting measurements of Higgs boson properties can be 
performed at a muon collider in the case where at least a hundred
inverse femtobarns of luminosity is available. Then the mass, width and spin 
of a SM-like Higgs boson can also be determined by operating either
a muon collider or a linear collider at the $Z\h$ production threshold where
the rate is sensitive to the Higgs mass\cite{Barger:1997pv}. 
With 100~fb$^{-1}$ of integrated luminosity, an error
of less than $100$~MeV can be achieved\cite{Barger:1997pv} for $\mh<150$~GeV. 
The shape of the $\ell^+\ell^-\to Z\h$ 
threshold cross section can also be used to determine the 
spin and to check the 
CP nature of the Higgs\cite{Miller:2001bi}. 

\section{Heavy Higgs Bosons}

In supersymmetric models there are multiple physical Higgs bosons. Often
the Higgs spectrum includes a SM-like Higgs boson with mass close to the 
$Z$ boson mass and some heavier Higgs bosons whose couplings are very 
much different than a SM particle of the same mass. For example, in the 
MSSM there is a light, neutral $\hl$ and two heavier neutral Higgs bosons,
$\hh$ and $\ha$. As one adjusts the parameters of the theory to make the
$\hh$ and $\ha$ heavier, the light Higgs boson $\hl$ becomes more and more
like the SM Higgs boson. It may very well be the case that after the initial
discovery of this SM-like Higgs boson the primary question will involve 
detecting deviations from the SM Higgs sector by a) measuring very 
precisely the SM-like Higgs boson properties, and/or b) directly discovering
additional Higgs bosons.

In the context of the MSSM, It is highly likely that the process 
$e^+e^-\to ZH$ used to find and study
the light Higgs state at a first generation LC 
will not be suitable for the heavier Higgs bosons, 
because in the decoupling limit the coupling of the Higgs to gauge bosons
is greatly suppressed (this is a corollary to the statement that the light
Higgs boson in Standard Model-like). There is a $250-500\gev$
range of heavy Higgs boson masses for which discovery is not possible 
via $\hh\ha$ pair production at a $\rts=500\gev$ LC. 
Further, the $\ha$ and $\hh$
cannot be detected in this mass range at either the LHC or LC 
for a wedge of moderate $\tanb$ values. (For large enough 
values of $\tanb$ the heavy Higgs bosons are expected to be observable
in $b\anti b \ha,b\anti b \hh$ production
at the LHC via their $\tau ^+\tau ^-$ decays and also at the LC.)
A linear collider operating in the $\gamma \gamma $ mode can produce 
Higgs bosons in the $s$-channel, and there have been a number of 
studies of such 
processes\cite{Jikia:1993di,Berger:1993tr,Dicus:1994ux,Gounaris:2000un,Muhlleitner:2001kw,Asner:2001ia,Berger:1992nr}. This requires that such an 
option exists, and the energy of the $\gamma \gamma $ system is not 
as sharply peaked at the center-of-mass energy as it is for the muon 
collider.

A muon collider can fill some, perhaps all of this moderate $\tanb$ wedge.
If $\tanb$ is large, the $\mupmum \hh$ and $\mupmum\ha$ couplings (proportional
to $\tanb$ times a SM-like value) are enhanced,
thereby leading to enhanced production rates in $\mupmum$ collisions.
These bosons can be discovered
via the radiative return mechanism\cite{Barger:1997jm}, 
and once a peak is found the 
machine energy can be set to $\mha$ or $\mhh$ and the muon collider becomes
a Higgs factory for the heavier Higgs bosons.
The resolution requirements for studying the 
heavy Higgs bosons in the 
$s$-channel are not as stringent as those for the light Higgs boson because
the heavier Higgs boson widths are generally much larger.
Since $R=0.1\%$ is sufficient, much higher luminosity
($L\sim 2-10~{\rm fb}^{-1}
/{\rm yr}$) would be possible as compared to that 
for $R=0.01\%-0.003\%$ as required for studying the $\hl$.
 
\begin{figure}
\centering\leavevmode
\includegraphics[width=5in]{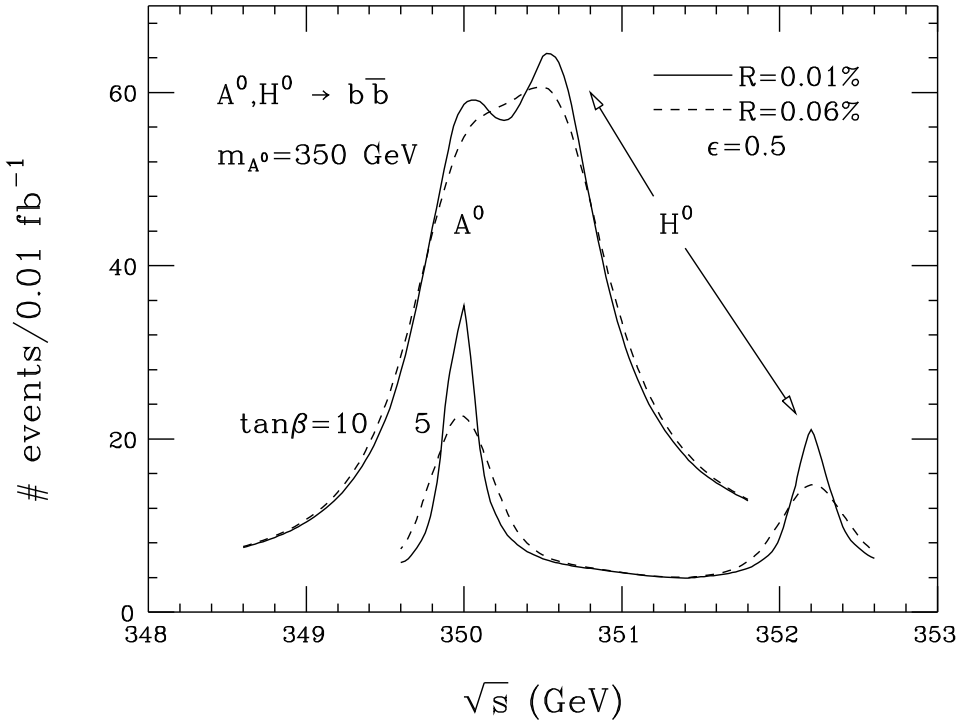}
\caption[Separation of $A$ and $H$ signals for $\tan\beta=5$ and $10$]
{Separation of $A$ and $H$ signals for $\tan\beta=5$ and $10$. From  
Ref.~\cite{Barger:1997jm}. \label{H0-A0-sep}}
\end{figure}

In the MSSM, the heavy Higgs bosons are largely degenerate, especially in the 
decoupling limit where they are heavy. 
In that case, a muon collider with sufficient energy resolution might be
the only possible means for separating out these states.
Examples showing the $\hh$ and $\ha$ resonances for $\tan \beta =5$ and $10$
are shown in Fig.~\ref{H0-A0-sep}. For the larger value of 
$\tan \beta$ the resonances are clearly overlapping. For the better energy 
resolution of $R=0.01\%$, the two distinct resonance peaks are still 
visible, but they are smeared out and merge into one broad
peak for $R=0.06\%$. 

Muon colliders excel at making precise measurements of Higgs boson masses
since they can exploit the $s$-channel production process.
This is reminiscent of the very accurate determination of the $Z$ boson mass
to just 2.2~MeV from the LEP measurements\cite{Groom:2000in}. Precise 
measurements of supersymmetric Higgs boson masses could provide a powerful
window on radiative corrections\cite{Berger:2001et}.
Supersymmetry together 
with gauge invariance in the MSSM implies the mass-squared sum rule
\begin{eqnarray}
&&m_{h^0}^2+m_{H^0}^2=m_{A^0}^2+m_Z^2+\Delta \;,
\end{eqnarray}
where $\Delta $ is a calculable radiative correction (the tree-level sum 
rule results from setting $\Delta =0$). This formula involves 
observables (masses) that can be precisely 
measured in the $s$-channel processes. Solving for the mass difference
\begin{eqnarray}
&&m_{A^0}-m_{H^0}={{m_{h^0}^2-m_Z^2-\Delta}\over {m_{A^0}+m_{H^0}}}\;,
\end{eqnarray} 
and one obtains a form that indicates in the decoupling limit, 
$m_{A^0}\to \infty$,
the mass difference between the heavy Higgs bosons becomes small.
As discussed in the previous section, the light Higgs mass $m_{h^0}$
can be measured to
less than an MeV in the $s$-channel. 
The masses of and the mass difference between
the heavy Higgs states $H^0$ and $A^0$ can also
be measured precisely by $s$-channel production. The ultimate
precision that can be obtained on the masses of the $H^0$ and $A^0$ depends 
strongly on the masses themselves and $\tan \beta$. But a reasonable 
expectation is that a scan through the resonances should be able to 
determine the masses and the mass-difference to some tens of 
MeV with just $0.1$~fb$^{-1}$ of integrated luminosity\cite{Berger:2001et}.
Altogether these mass measurements yield a value for the radiative 
correction $\Delta$ to a precision of order $10$~GeV$^2$. Since the 
typical size of $\Delta$ is of order $10^4$~GeV$^2$, this constitutes a 
measurement of roughly one part in $10^3$.
The quantity $\Delta$ is calculable in terms of the self-energy diagrams 
of the Higgs bosons\cite{Berger:1990hg}, and a comparison between 
the measured value and the theoretical prediction yields a test of 
radiative corrections in the MSSM. Further progress in the theoretical
calculation of $\Delta$ would be needed to fully exploit the expected
precision of the experimental measurements.

\section{Concluding Remarks}

Recent experimental results hint that a muon collider may play a crucial
role in studying the next generation of physics signals. 
There is the evidence from 
LEP\cite{Barate:2000ts,Abreu:2000fw,Acciarri:2000ke,Abbiendi:2000ac,Okpara:2001jf}  
for a Higgs boson near $m_H\simeq 115$~GeV. This $\gsim 2\sigma$ signal is not
definitive, but it has been taken very seriously 
since it is consistent with the current precision 
electroweak data and fits well with a supersymmetric interpretation. 
A Higgs boson with such a mass
is in the optimal range for study at a Higgs factory.
Such a Higgs boson sits comfortably above the $Z$-pole where there
is a large background from 
$Z$ decay to $b\overline{b}$, and a $115\gev$ mass 
is sufficiently below the $WW^\star$ threshold  that 
the decay width remains small and the ability of the muon collider to
achieve a very narrow beam energy spread can be exploited.

In the MSSM such a Higgs boson mass of $115$~GeV is near
the theoretical upper limit of $m_{H^0}<130$~GeV, and would indicate
a value of the supersymmetry parameter $\tan \beta$
substantially above 1 (assuming stop masses $\lsim 1\tev$). 
This is consistent with recent evidence for non-SM contributions 
to the anomalous magnetic moment of the 
muon\cite{Brown:2001mg} which also can be explained in the MSSM with a 
moderately large value of $\tan \beta$. If these early indications prevail,
and we are left with a supersymmetric Higgs sector with large $\tan \beta$, 
then it is likely that the heavy Higgs $\hh$ and $\ha$ will not be 
observable at the LHC or a LC. The detection of these Higgs bosons could be
accomplished in the $s$-channel at a muon collider, and some precision
tests involving the Higgs boson masses can be performed to check radiative
corrections in the Higgs sector.
More generally, the muon collider has the potential to find and study
Higgs bosons that exist in more general models than the MSSM with extended
Higgs sectors. In this more general context, the muon collider offers the 
possibility of studying the CP nature of the Higgs bosons that are found.

Finally the muon collider program encompasses much more than physics of Higgs 
factories described here. Interesting physics can be envisioned at all stages
of the development of muon colliders: from neutrino factories to Higgs 
factories to even higher energies.

%
%

%
%

\begin{acknowledgments}
This work was supported in part by the U.S.
Department of Energy under Grant No.~DE-FG02-91ER40661.
\end{acknowledgments}

\bibliography{P1_berger_0717}

\begin{thebibliography}{10}
\providecommand*{\bibinfo}[2]{#2}
\providecommand*{\eprint}[1]{#1}
\providecommand*{\url}[1]{#1}
\bibitem{e1group}
\bibinfo{author}{V.~Barger}, \bibinfo{author}{M.~S. Berger},
  \bibinfo{author}{J.~F. Gunion}, and \bibinfo{author}{T.~Han},
  \bibinfo{title}{\emph{Physics of higgs factories}} (\bibinfo{date}{2001}),
  \eprint{arXiv:hep-ph/0110340}.
\bibitem{Ankenbrandt:1999as}
\bibinfo{author}{C.~M. Ankenbrandt} \emph{et~al.}, \bibinfo{journal}{Phys. Rev.
  ST Accel. Beams} \bibinfo{volume}{\textbf{2}}, \bibinfo{pages}{081001}
  (\bibinfo{date}{1999}), \eprint{arXiv:physics/9901022}.
\bibitem{Barger:1997jm}
\bibinfo{author}{V.~Barger}, \bibinfo{author}{M.~S. Berger},
  \bibinfo{author}{J.~F. Gunion}, and \bibinfo{author}{T.~Han},
  \bibinfo{journal}{Phys. Rept.} \bibinfo{volume}{\textbf{286}},
  \bibinfo{pages}{1} (\bibinfo{date}{1997}), \eprint{arXiv:hep-ph/9602415}.
\bibitem{Barger:1995hr}
\bibinfo{author}{V.~Barger}, \bibinfo{author}{M.~S. Berger},
  \bibinfo{author}{J.~F. Gunion}, and \bibinfo{author}{T.~Han},
  \bibinfo{journal}{Phys. Rev. Lett.} \bibinfo{volume}{\textbf{75}},
  \bibinfo{pages}{1462} (\bibinfo{date}{1995}), \eprint{arXiv:hep-ph/9504330}.
\bibitem{Ayres:1999ug}
\bibinfo{author}{D.~Ayres} \emph{et~al.} (\bibinfo{collaboration}{Neutrino
  Factory and Muon Collider}), \bibinfo{title}{\emph{Expression of interest for
  r \& d towards a neutrino factory based on a storage ring and a muon
  collider}} (\bibinfo{date}{1999}), \eprint{arXiv:physics/9911009}.
\bibitem{Holtkamp:2000xn}
\bibinfo{author}{N.~Holtkamp} \emph{et~al.}, \bibinfo{title}{\emph{A
  feasibility study of a neutrino source based on a muon storage ring}},
  slac-reprint-2000-054.
\bibitem{Raja:1998ip}
\bibinfo{author}{R.~Raja} and \bibinfo{author}{A.~Tollestrup},
  \bibinfo{journal}{Phys. Rev.} \bibinfo{volume}{\textbf{D58}},
  \bibinfo{pages}{013005} (\bibinfo{date}{1998}),
  \eprint{arXiv:hep-ex/9801004}.
\bibitem{Battaglia:2000jb}
\bibinfo{author}{M.~Battaglia} and \bibinfo{author}{K.~Desch}
  (\bibinfo{date}{2000}), \eprint{arXiv:hep-ph/0101165}.
\bibitem{Barger:1997pv}
\bibinfo{author}{V.~Barger}, \bibinfo{author}{M.~S. Berger},
  \bibinfo{author}{J.~F. Gunion}, and \bibinfo{author}{T.~Han},
  \bibinfo{journal}{Phys. Rev. Lett.} \bibinfo{volume}{\textbf{78}},
  \bibinfo{pages}{3991} (\bibinfo{date}{1997}), \eprint{arXiv:hep-ph/9612279}.
\bibitem{Miller:2001bi}
\bibinfo{author}{D.~J. Miller}, \bibinfo{author}{S.~Y. Choi},
  \bibinfo{author}{B.~Eberle}, \bibinfo{author}{M.~M. Muhlleitner}, and
  \bibinfo{author}{P.~M. Zerwas}, \bibinfo{journal}{Phys. Lett.}
  \bibinfo{volume}{\textbf{B505}}, \bibinfo{pages}{149} (\bibinfo{date}{2001}),
  \eprint{arXiv:hep-ph/0102023}.
\bibitem{Jikia:1993di}
\bibinfo{author}{G.~Jikia}, \bibinfo{journal}{Nucl. Phys.}
  \bibinfo{volume}{\textbf{B405}}, \bibinfo{pages}{24} (\bibinfo{date}{1993}).
\bibitem{Berger:1993tr}
\bibinfo{author}{M.~S. Berger}, \bibinfo{journal}{Phys. Rev.}
  \bibinfo{volume}{\textbf{D48}}, \bibinfo{pages}{5121} (\bibinfo{date}{1993}),
  \eprint{arXiv:hep-ph/9307259}.
\bibitem{Dicus:1994ux}
\bibinfo{author}{D.~A. Dicus} and \bibinfo{author}{C.~Kao},
  \bibinfo{journal}{Phys. Rev.} \bibinfo{volume}{\textbf{D49}},
  \bibinfo{pages}{1265} (\bibinfo{date}{1994}), \eprint{arXiv:hep-ph/9308330}.
\bibitem{Gounaris:2000un}
\bibinfo{author}{G.~J. Gounaris}, \bibinfo{author}{P.~I. Porfyriadis}, and
  \bibinfo{author}{F.~M. Renard}, \bibinfo{journal}{Eur. Phys. J.}
  \bibinfo{volume}{\textbf{C19}}, \bibinfo{pages}{57} (\bibinfo{date}{2001}),
  \eprint{arXiv:hep-ph/0010006}.
\bibitem{Berger:1992nr}
\bibinfo{author}{M.~S. Berger} (\bibinfo{date}{1992}),
  \eprint{arXiv:hep-ph/9207275}.
\bibitem{Muhlleitner:2001kw}
\bibinfo{author}{M.~M. Muhlleitner}, \bibinfo{author}{M.~Kramer},
  \bibinfo{author}{M.~Spira}, and \bibinfo{author}{P.~M. Zerwas},
  \bibinfo{journal}{Phys. Lett.} \bibinfo{volume}{\textbf{B508}},
  \bibinfo{pages}{311} (\bibinfo{date}{2001}), \eprint{hep-ph/0101083}.
\bibitem{Asner:2001ia}
\bibinfo{author}{D.~M. Asner}, \bibinfo{author}{J.~B. Gronberg}, and
  \bibinfo{author}{J.~F. Gunion} (\bibinfo{date}{2001}),
  \eprint{arXiv:hep-ph/0110320}.
\bibitem{Groom:2000in}
\bibinfo{author}{D.~E. Groom} \emph{et~al.} (\bibinfo{collaboration}{Particle
  Data Group}), \bibinfo{journal}{Eur. Phys. J.}
  \bibinfo{volume}{\textbf{C15}}, \bibinfo{pages}{1} (\bibinfo{date}{2000}).
\bibitem{Berger:2001et}
\bibinfo{author}{M.~S. Berger}, \bibinfo{journal}{Phys. Rev. Lett.}
  \bibinfo{volume}{\textbf{87}}, \bibinfo{pages}{131801}
  (\bibinfo{date}{2001}), \eprint{arXiv:hep-ph/0105128}.
\bibitem{Berger:1990hg}
\bibinfo{author}{M.~S. Berger}, \bibinfo{journal}{Phys. Rev.}
  \bibinfo{volume}{\textbf{D41}}, \bibinfo{pages}{225} (\bibinfo{date}{1990}).
\bibitem{Barate:2000ts}
\bibinfo{author}{R.~Barate} \emph{et~al.} (\bibinfo{collaboration}{ALEPH}),
  \bibinfo{journal}{Phys. Lett.} \bibinfo{volume}{\textbf{B495}},
  \bibinfo{pages}{1} (\bibinfo{date}{2000}), \eprint{arXiv:hep-ex/0011045}.
\bibitem{Abreu:2000fw}
\bibinfo{author}{P.~Abreu} \emph{et~al.} (\bibinfo{collaboration}{DELPHI}),
  \bibinfo{journal}{Phys. Lett.} \bibinfo{volume}{\textbf{B499}},
  \bibinfo{pages}{23} (\bibinfo{date}{2001}), \eprint{arXiv:hep-ex/0102036}.
\bibitem{Acciarri:2000ke}
\bibinfo{author}{M.~Acciarri} \emph{et~al.} (\bibinfo{collaboration}{L3}),
  \bibinfo{journal}{Phys. Lett.} \bibinfo{volume}{\textbf{B495}},
  \bibinfo{pages}{18} (\bibinfo{date}{2000}), \eprint{arXiv:hep-ex/0011043}.
\bibitem{Abbiendi:2000ac}
\bibinfo{author}{G.~Abbiendi} \emph{et~al.} (\bibinfo{collaboration}{OPAL}),
  \bibinfo{journal}{Phys. Lett.} \bibinfo{volume}{\textbf{B499}},
  \bibinfo{pages}{38} (\bibinfo{date}{2001}), \eprint{arXiv:hep-ex/0101014}.
\bibitem{Okpara:2001jf}
\bibinfo{author}{A.~N. Okpara} (\bibinfo{date}{2001}),
  \eprint{arXiv:hep-ph/0105151}.
\bibitem{Brown:2001mg}
\bibinfo{author}{H.~N. Brown} \emph{et~al.} (\bibinfo{collaboration}{Muon
  g-2}), \bibinfo{journal}{Phys. Rev. Lett.} \bibinfo{volume}{\textbf{86}},
  \bibinfo{pages}{2227} (\bibinfo{date}{2001}), \eprint{arXiv:hep-ex/0102017}.

\end{thebibliography}

\end{document}